\documentclass[a4paper]{article}
\usepackage[a4paper,top=3cm,bottom=2cm,left=3cm,right=3cm,marginparwidth=1.75cm]{geometry}
\usepackage{amsmath, amsthm, amsfonts}
\usepackage{bm}
\usepackage{graphicx}
\usepackage{slashed} 
\usepackage[colorinlistoftodos]{todonotes}
\usepackage[colorlinks=true, allcolors=blue]{hyperref}
\usepackage[sorting=none, backend=biber]{biblatex}
\bibliography{Bibliography}

\title{Neutrino electromagnetic properties in a Leptoquark model}
\author{R. S\'anchez-V\'elez\footnote{corresponding author ricardo.sanchez.v@cinvestav.mx} \\ \\
              \small{\textit{Departamento de F\'isica, Centro de Investigaci\'on y de Estudios   
                        Avanzados del IPN}},\\ 
              \small{\textit{Apartado Postal 14-740 07000 Ciudad de M\'exico, Mexico}} 
            }
\date{}
\begin{document}

\maketitle

\begin{abstract}
The neutrino magnetic dipole moment is studied in a simple extension of the SM augmented with a scalar Leptoquark with quantum numbers $(\bar{\boldsymbol{3}},\boldsymbol{1},1/3)$. In order to reach a sizeable neutrino magnetic moment, we include in the framework the contribution of right-handed neutrinos and analytical result for the one-loop contribution of a scalar LQ is presented. The allowed parameter space is found through recent restrictions on the muon $(g-2)$, the $\tau \to \mu \gamma$ and the $B_c \to \tau^- \bar \nu$ decay. It is shown that exist a region of the parameter space where the recent anomaly reported by the XENON collaboration can be accommodated. 
\end{abstract}

\section{Introduction}

In the standard model neutrinos are massless and have no electric or magnetic dipole moment, however from various experiments we know that neutrinos have mass. Since neutrinos do not carry electric charge, they can participate in electromagnetic interactions by coupling with photons only through quantum corrections. Then, in the minimal extension of the Standard Model with three right-handed neutrinos, the magnetic moment $\mu_\nu$ can be calculated at one-loop level \cite{Fujikawa:1980yx}

\begin{equation}\label{eq:NeutrinoMagnetic}
\mu_\nu=\frac{3 m_e G_F}{4\sqrt{2} \pi^2}m_\nu \mu_B\approx 3.2\times 10^{-19}\left(\frac{\mu_\nu}{eV}\right)\mu_B,
\end{equation}

where $m_e$ is the electron mass, $G_F$ the Fermi coupling constant and $\mu_B$ the Bohr magneton as a conventional unit. On the experimental side, laboratory limits are determined via neutrino-electron scattering at low energies, with $\mu_\nu < 1.5\times 10^{-10} \mu_\text{B}$ \cite{Super-Kamiokande:2004wqk} and $\mu_\nu < .7\times 10^{-10} \mu_\text{B}$ \cite{TEXONO:2006xds}, obtained from solar and reactor experiments, respectively. Recently, the XENON Collaboration reported an excess of events at low energies between 2 and 3 keV in their dual-phase noble gas detector \cite{XENON:2020rca}. This experiment was designed to find WIMP particles, which are dark matter candidates with masses of the order of $6$ GeV$/c^2$ and 1 TeV$/c^2$. Thanks to a new estimate of the sensitivity in electronic recovery regions below 20 keV, new data between 1 and 30 keV have been published, where 285 events with energy between 1 and 7 KeV were observed, when $232\pm 15$ events were expected. Among other explanations, this excess can be interpreted as a signal from solar neutrinos with an enhanced neutrino magnetic moment. With this idea, several studies have been carried  to explain the excess of events, for example scenarios with dark photons \cite{An:2020bxd}, dark matter decay \cite{Farzan:2020llg}, axion models \cite{Bloch:2020uzh} or a model based on $SU(2)_\text{H}$ horizontal symmetry \cite{Babu:2020ivd} have been employed. Therefore 	the analysis of the neutrino magnetic moment is a great opportunity to find evidence beyond the Standard Model. \\ 
\\
In this work, we consider the possibility of using scalar Leptoquarks interactions to generate a non-zero neutrino magnetic moment that may lie within the experimental reach. In addition, we are interested in regions of the LQ parameter space where the excess recently reported by the XENON collaboration might fit in with other constrains coming from the flavor physics. Leptoquarks are hypothetical color-triplet bosons that appear naturally in the context of Grand Unified Theories \cite{Pati:1974yy,Georgi:1974sy,Fritzsch:1974nn}, where strongly non-interacting leptons are accommodated into the same multiplets as quarks. There are also other well established theories that predict the existence of Leptoquarks, such as technicolor \cite{Ellis:1980hz,Farhi:1980xs,Hill:2002ap}, $R$-parity violating supersymmetric models \cite{Barbier:2004ez}, models with composite fermions \cite{Schrempp:1984nj,Buchmuller:1985nn,Gripaios:2009dq}, etc. These non-Standard Model particles can be scalar or vector nature and have the interesting property of turning leptons into quarks and vice versa. The physics of Leptoquarks can be studied systematically based on their representation under the Standard Model gauge group $SU(3)\times SU(2)\times U(1)$ \cite{Buchmuller:1986zs}, where ten different Leptoquark states can emerge if one assumes the SM with purely left-handed neutrinos. However, more LQ states arise if electrically neutral states that play the role of right-handed neutrinos are added to the SM particle spectrum. Recently, the analysis of Leptoquark has become more relevant due to the possibility of describing certain anomalies. The discrepancies that have attracted more attention are $R_{D^{(*)}}$ \cite{Mandal:2018kau}, $R_k$ \cite{Becirevic:2017jtw}, the so called $B$ anomalies \cite{Becirevic:2016yqi} and the anomalous magnetic moment of the muon \cite{Bigaran:2020jil}. Their existence might also give a hint on why there are exactly three generations of matter or why there are the same number of species  of quark and lepton, the result of which is the fact that the currents associated with the SM gauge symmetries are non-anomalous.  Following the above understanding, Leptoquarks are currently among the  most important contributions of NP. However, despite the immense experimental effort, they have not been directly observed yet. In the past, different studies have contemplate the LQ contribution to $\nu$MM, namely the authors of Ref. \cite{Chua:1998yk} employed vector Leptoquarks to generate a $\mu_\nu$ of the order of  $10^{-10}$($10^{-12}$)  for third (second) generation LQs. Furthermore, the scalar LQs phenomenology was studied in Ref. \cite{Povarov:2007} within the minimal model with four-color symmetry where constraints on the LQ mass were predetermined from the astrophysical data on the neutrino magnetic moment.  On the other hand, it turns out that masses and couplings of vector LQs remain severely constrained by the experimental data, which is why scalar LQs analysis has been favoured in the literature. In this paper we study the neutrino magnetic moment in a framework where the SM is augmented with a scalar LQ singlet under $SU(2)$ with hypercharge $1/3$, which is usually denoted as $S_1$ in the literature. As we consider non-chiral LQ couplings to neutrinos and quarks, the neutrino magnetic moment is enhanced by the quark masses running inside the loop, especially with heavy quarks. In addition, we also apply other well-studied LQ induced processes, such as the muon anomalous magnetic moment, the lepton flavor-violating decays and the $B$ meson pure leptonic decays. This processes constraint the LQ couplings and its mass. \\ 
\\
The organization of the paper goes as follows. In Sect. \ref{sec:LQmodel} we briefly discuss the framework of the LQ model we are interested in. Section \ref{sec:NeutrinoMagnetic} is devoted to present the general calculation of the neutrino magnetic moment induced by the scalar LQ $S_1$. In Sect. \ref{sec:Constraints} we present a discussion on the constraints on the LQ couplings from experimental data, followed by the numerical analysis of the neutrino magnetic moment in Sect. \ref{sec:Analysis}. Finally, the conclusions and outlook are presented in Sect. \ref{sec:Summary}. 

\section{A simple model of scalar LQ}\label{sec:LQmodel}

The LQ phenomenology is usually studied via a model independent approach through an effective lagrangian, which allow us to focus on the low-energy LQ interaction ignoring (without  lost of generality) the complex framework of the ultraviolet completion. The most general lagrangian of dimension four with effective interactions invariant under $SU(3)_c \times SU(2)_L \times U(1)_Y$ for both scalar and vector LQs was first presented in \cite{Buchmuller:1986zs} and for a recent review we recommend the Ref. \cite{Dorsner:2016wpm}. Namely, we are interested in the contribution of the singlet scalar Leptoquark $S_1$  as an explanation for the large neutrino magnetic moment needed to explain the recent anomaly reported by the XENON collaboration. The new scalar $S_1$ transforms as $(\bar{\boldsymbol{3}},\boldsymbol{1},1/3)$ under the standard model gauge group, and its couplings to fermions are described by the lagrangian 

\begin{align}\label{yukawa}
\mathcal L &= y^L_{ij} \bar Q_{iL}^{c} i\sigma_2 L_{jL} S_1 + y^R_{ij}\bar u^c_{iR} l_{jR} S_1 + \tilde{y}^{R}_{ij} \bar d_{iR}^c \nu_{Rj} S_1 + \text{h.c} ,
\end{align}

where $Q^c_{iL}$ and $L_{jL}$ denoted the left-handed quark and lepton doublet with flavor indices $i,j$. The fields $u_R$, $l_R$ are the right-handed up-type quark and charge lepton singlets respectively. The superscript $c$ in the fermion fields stands for the charge conjugation field defined as $\Psi^c$ where $\Psi^c= C\bar \Psi^T$ and $\bar \Psi^c=-\Psi^T C^{-1}$ with $C$ the charge conjugation matrix. $y_{ij}^H$ represents the Yukawa coupling of $S_1$ with a charge conjugate quark from generation $i$ and a lepton of chirality $H$ from generation $j$. Note that we are considering the possibility that $S_1$ couples to the right-handed neutrinos. It is worth mentioning that the Leptoquark $S_1$ has lepton and baryon number violating interactions that can mediate the proton decay at the tree level. However, such operators can be eliminated by a discrete symmetry in which opposite parity is assigned to the SM leptons and Leptoquarks. The phenomenology of the LQ $S_1$ augmented with right-handed neutrinos has been recently  analysed in \cite{Dorsner:2019vgp} where the authors assume couplings with the first generation fermions only. Rotating the Lagrangian from the weak to the mass basis for quarks and leptons, the interaction terms take the form 

\begin{align}
\mathcal L &= - y^L_{i j} \bar d_{i L}^c \nu_{j L} S_1 + V_{i k} y^L_{i j} \bar u_{k L}^c l_{j L} S_1 + y_{k j}^R \bar u_{k j}^c l_{j R} S_1 + \tilde y^R_{i j } \bar d^c_{i R} \nu_{R j} S_1 + \text{h.c}. 
\end{align}
where we assumed that $S_1$ is aligned to the down-type quark basis. Note that the $S_1$ couplings with up-type quarks and charge leptons are fixed by the CKM mixing matrix. Besides of the Yukawa couplings, we also required the Leptoquark coupling to the photon, whose Feynman rule can be directly extracted from the LQ kinetic term

\begin{equation}
\mathcal{L}=\frac{1}{2}(D^\mu S_1)^\dagger D^\mu S_1,
\end{equation}
where the $SU(2)_L\times U(1)_Y$ covariant derivate is given by 
\begin{equation}
D_\mu S_1=\left(\partial_\mu+ig\frac{\tau^i}{2}W^i+ig^\prime \frac{1}{3}B^\mu\right)S_1.
\end{equation}

Then, the corresponding Feynman rule can be written as follows 

\begin{equation}
\mathcal{L} \supset \frac{i e}{3}S_{1}\overleftrightarrow{\partial_\mu} S_{1}^* A^\mu. 
\end{equation}

The complete lagrangian, including all the LQs interactions and the corresponding Feynman rules is presented in Ref. \cite{Crivellin:2021ejk}, that can be used for an automated analysis of Leptoquarks. 

\section{Leptoquark contribution to the neutrino magnetic moment}\label{sec:NeutrinoMagnetic}

\begin{figure}[hbt!]
\centering
\includegraphics[width=11cm]{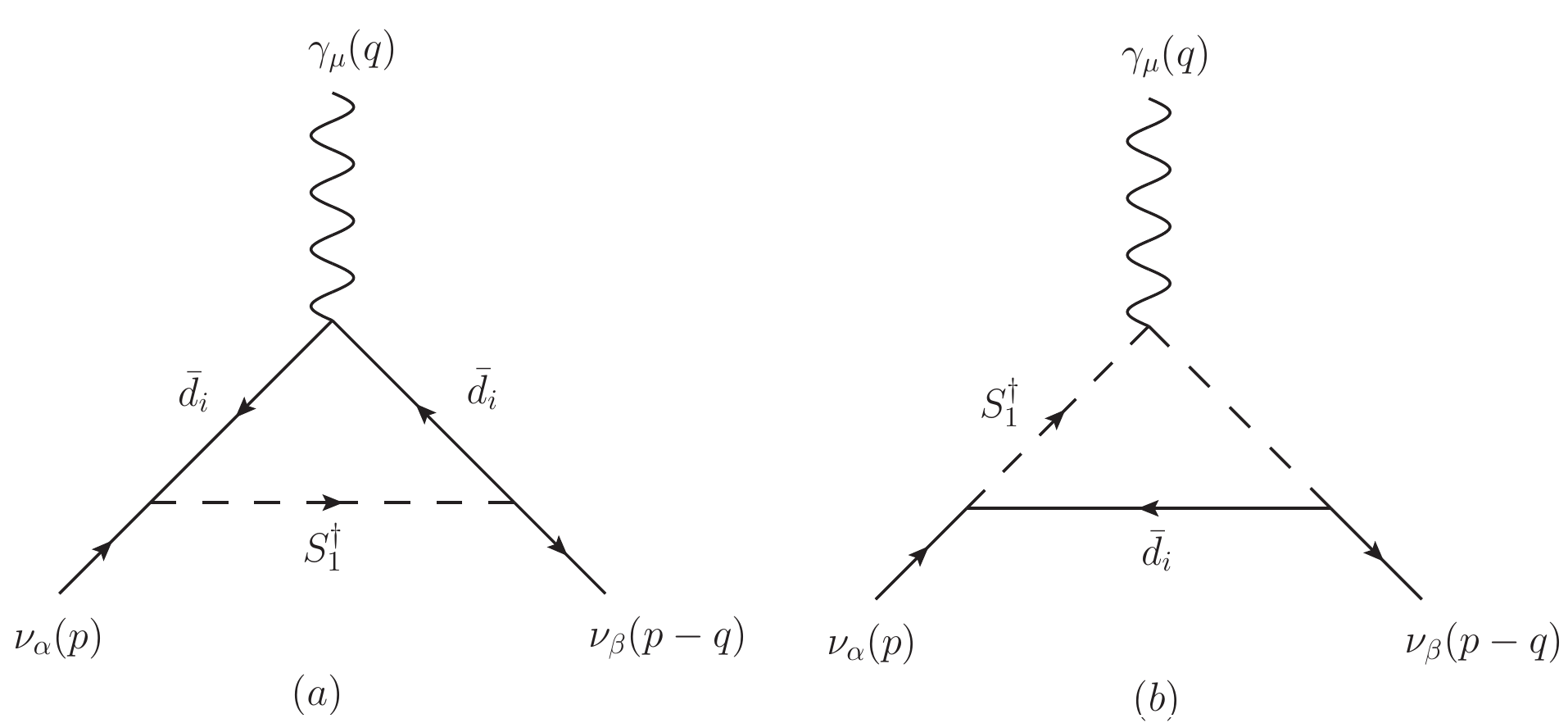}
\caption{One loop diagrams representing the scalar Leptoquark $S_1$ contribution to the neutrino magnetic dipole moment.}\label{Fig:FeynmanDiag1} 
\end{figure}

The magnetic moment of the neutrino, induced by the scalar LQ $S_1$, can be derived at one loop level from the Feynman diagrams shown in Fig. \ref{Fig:FeynmanDiag1}, where $\nu_\alpha=\nu_e,\nu_\mu,\nu_\tau$ and $\nu_\beta $ refers to a different neutrino flavour (the notation for the four-momenta of the external particles is also shown).

By using the Feynman rules given in the previous section we obtain the invariant amplitude for the Feynman diagrams (a)-(b) in Fig. \ref{Fig:FeynmanDiag1}

\begin{align}
\mathcal{M}^{(a)} &= i e Q_{\bar{d}} N_c \epsilon_\mu^*(q)  \bar u(p - q) \left\{ \int \frac{d^Dk}{(2\pi)^D}( -y^{L*}_{i\beta}     
 P_R + \tilde y^{R *}_{i \beta} P_L) \frac{ (\slashed{p} - \slashed{k} - \slashed{q} + m_{d_i} ) }{ (p - k - q)^2 - m_{d_i}^2 } \gamma^\mu \right. \nonumber \\ 
 	& \left. \times \frac{ ( \slashed{p} - \slashed{k} + m_{d_i} ) }{ (p - k)^2 - m_{d_i}^2}  ( -y^L_{i\alpha} P_L + y^{\prime R}_{i \alpha} P_R ) \frac{1}{ k ^2 - m_{S_1}^2} \right \} u(p),\\
\mathcal{M}^{(b)} &= - i e Q_{S_1} N_c \epsilon_\mu^*(q) \bar u(p - q) \left \{ \int \frac{d^D k}{ (2\pi)^D }  ( 2 p - 2 k -q )^\mu ( - y^{L *}_{i \beta} P_R + \tilde y^{R *}_{i \beta} P_L )\frac{ \slashed{k} + m_{d_i} }{ k^2 - m_{d_i}^2 } \right. \nonumber \\
	& \left. \times ( - y^L_{i \alpha} P_L + \tilde y^{R }_{i \alpha } P_R ) \frac{1}{ (p - k)^2 - m_{S_1}^2}\frac{1}{ (p - q - k)^2 - m_{S_1}^2 } \right \} u(p),
\end{align}

where $P_L$ and $P_R$ are the projector operators and $Q_{\bar d} = 1/3 $ ($Q_{S_1} = 1/3 $) is the quark $\bar d$ (scalar Leptoquark $S_1$) electric charge in units of the elemental electric charge. The masses of the scalar Leptoquark and the down-quark particles are $m_{S_1}$ and $ m_{d_i}$ respectively. As we can see from Fig. \ref{Fig:FeynmanDiag1} there are two fermion-flow clashing into a vertex due to the Feynman rules for such a LQ of Majorana-like type that requires a special treatment. We have followed the approach of Ref. \cite{Crivellin:2021ejk}  for the interactions involving charge-conjugate SM fermions. We employ the Feynman parametrization technique to evaluate the amplitudes of the Feynman diagrams, which allow us to write the generalized form of a neutrino electromagnetic vertex function     

\begin{equation}
\mathcal M = \bar u (p-q) \Bigl ( (\gamma^\mu - q^\mu \slashed{q} /q^2) (f_1^{\alpha \beta }(q^2) +f^2_{\alpha \beta} (q^2) q^2 \gamma^5 ) -i \sigma^{\mu \nu} q_\nu (f^3_{\alpha \beta} (q^2) + i f^4_{\alpha \beta} (q^2) \gamma^5) \Bigr) u(p).
\end{equation}

In the case of coupling with a real photon we define the magnetic moment of the neutrino as the magnetic form factor $\mu_{\nu_\alpha \nu_\beta} = f^3_{\alpha \beta} (0)$ taken at $q^2=0$. Then, the contribution of the Leptoquark $S_1$ and quarks-down to the neutrino magnetic moment at one loop-level can be express in the following form 

\begin{align}\label{neutrinomoment}
\mu_{\nu_{\alpha \beta}} &= - \frac{N_c m_e}{16 \pi^2 m_{S_1}^2} \mu_B \sum_i \left( (y^L_{i \alpha} y^{L*}_{i \beta} + y^{\prime R}_{i \alpha} y^{\prime R*}_{i \beta} )\frac{m_\alpha + m_\beta }{2} \Bigl [ Q_{\bar d} G_1 (a_i) - Q_{S_1} G_2 (a_i) \Bigr ] \right. \nonumber\\
&\left. + ( y^L_{i \alpha} \tilde y^{R *}_{i \beta } + y^{L *}_{i \beta} \tilde y^{R}_{i\alpha } )\frac{m_{d_i}}{2} \Bigl [ Q_{\bar d} G_3(a_i) - Q_{S_1} G_4(a_i) \Bigr ] \right),
\end{align}

with $m_{\alpha , \beta}$ the neutrino masses with flavor $ \alpha , \beta $. We also define the scale variable $a_i=m_{d_i}^2/m_{S_1}^2$. The different functions $G_i$ are given by 

\begin{align*}
G_1(a) & = \frac{2 + 3 a - 6 a^2 + a^3 + 6 \; \text{ln}(a)}{12 (1 - a)^4},\\
G_2(a) & = \frac{ -1 +6 a - 3 a^2 - 2 a^3 +6 a^2  \text{ln}(a) }{12(1 - a)^4},\\
G_3(a) & = \frac{ 3 - 4 a + a^2 + 2 \; \text{ln}(a)  }{2 (1 - a)^3}, \\
G_4(a) & = \frac{ 1 - a^2 + 2 a \; \text{ln}(a) }{ 2(1 - a)^3 }.
\end{align*}

As expected for a non-chiral Leptoquark, besides of the term proportional to the neutrino mass, the neutrino magnetic moment is also proportional to the internal quark-down mass, where the quark-bottom yields the main contribution. 

\section{ Constraints on the parameter space of the scalar LQ model}\label{sec:Constraints}

In this section we provide the treatment over the space parameter for the LQ model described in section \ref{sec:LQmodel}. We will use the discrepancy of the muon anomalous magnetic dipole moment and the experimental bound on the lepton flavor violating decay $ \tau \to \mu \gamma $ to restrict the Yukawa couplings $ y^{L,R}_{i,j} $. In addition, we discuss the latest limits on the Leptoquark mass imposed by accelerator experiments.

\subsection{Constraints on the LQ mass}
Because the discovery of Leptoquarks may be a clear indication of physics beyond the SM, several searches for LQs have been undertaken in previous experiments, such as HERA and the Tevatron \cite{Acosta:1999ws}, as well as in the recent colliders. For example, the collider phenomenology was recently analysed by the ATLAS collaboration at $\sqrt s = 13 $ TeV and an integrated luminosity of $3.2\, \text{fb}^{-1}$, where the constraints of the couplings of LQ to third and second generation fermions as well as the mass of the LQ were obtained \cite{ATLAS:2016wab}. Once the LQ is allowed to decay into $u e$ ($c \mu$) with $100 \%$ of the branching fraction, the mass of the LQ is constrained to $m_{LQ} > 1 \, (1.2)$ TeV. CMS collaboration has also searched for scalar Leptoquarks with electrical charges $-1/3$ and $-5/3$ through proton-proton collisions at $\sqrt s = 13$ TeV and an integrated luminosity of $35.9 \, \text{fb}^{-1}$. Considering a third generation LQ that only decay to a quark top and a tau lepton, masses below $900$ GeV are excluded at $95 \, \%$ confidence level \cite{CMS:2018svy}. A less restricted bound on the LQ mass of $m_{LQ} > 660 $ GeV was derived by previous data from the search of single Leptoquark production at $\sqrt s = 8$ TeV of the CMS experiment \cite{CMS:2015xzc}. Then, we will assume the less stringent bound $m_{S_1} > 900$ GeV in our analysis, since we are considering a LQ that couples to both second and third-generation fermions. \\

\subsection{Anomalous magnetic moment of the muon and radiative lepton decays}\label{subsec:SpaceParameter}

The Leptoquark scalar model described in Sec. \ref{sec:LQmodel} can induced important contributions to certain observables that have already been accurately measured. Among the most restrictive processes are the discrepancy of the muon anomalous magnetic dipole moment and the lepton flavor violating decays $l_i \to l_j \gamma$, which can be affected by the presence of scalar Leptoquarks. Although the contribution of scalar LQ to both processes has been extensively studied in the literature, in this work we reproduce the relevant calculations to derive the available space parameter. The contribution of $S_1$ to the LFV decay $l_i \to l_j\gamma$ arises at the one-loop level by Feynman diagrams similar to the one shown in Fig. \ref{Fig:FeynmanDiag1} with the following substitutions in the external leptons $\nu_{ \alpha,\beta} \to l_{i , j}$ and the internal quarks $ \bar d_i \to \bar u_i $. Moreover, there are contributions from  reducible diagrams, however they only give contributions to the monopole terms which cancel out with those arising from the irreducible diagrams due to gauge invariance. Then, the decay amplitude $l_i^- \to l_j^- \gamma$ can be written as follows

\begin{equation}\label{Eq:Amplitude2}
\mathcal{M} (l_i^- \to l_j^- \gamma) = - \frac{i e}{16 \pi^2} \epsilon_\mu^* (q) \bar u (p - q) \left ( A_L P_L + A_R P_R  \right ) \sigma^{\mu \nu} q_\nu u (p),
\end{equation}

where the form factors are given by

\begin{align}
A_L &= \sum_{k=1}^3 \frac{N_c }{m_{S_1}^2} \Bigl [ m_{k} \tilde y^{L}_{k i} y^{R*}_{k j} \bigl( Q_S  G_4 (b_k) - Q_{\bar u} G_3 (b_k) \bigr) \nonumber \\ 
	  & - \bigl(m_i y^R_{k i} y^{R*}_{k j} + m_j  y^{\prime L}_{k i} \tilde y^{L*}_{k j} \bigr) \bigl( Q_S G_2 (b_k) - Q_{\bar u} G_1 (b_k) \bigr) \Bigr ], \\
A_R &= A_L ( y^{\prime L}_{k l} \leftrightarrow y^R_{k l}), \quad l = i , j , 
\end{align}

where $Q_{\bar u} = - 2/3$ is the quark $\bar u$ electric charge and $b_k = m_{k} ^2/ m_{S_1}^2$ with $ m_{k} = (m_u, m_c, m_t) $. The Yukawa left-handed coupling is given by $y^{\prime L}_{k l} = V_{a k} y^L_{a l} $. Then, after averaging (summing) over polarizations of the initial (final) fermion and gauge boson, we use the respective two-body decay width formula to write down the branching ratio of $l_i^- \to l_j^- \gamma$ 

\begin{equation}
\mathcal{B} (l_i^- \to l_j^- \gamma) = \frac{\alpha_{ \text{em} } (m_i^2 - m_j^2)^3}{4 (4 \pi)^4 m_i^3 \Gamma_i} \left ( |A_L|^2 + |A_R|^2 \right ),
\end{equation}

with $\alpha_\text{em} = e^2/(4 \pi)$ and $\Gamma_i$ being the fine-structure constant and the total decay width of the charge lepton $l_i^-$ respectively. From Eq. \ref{Eq:Amplitude2} we can subtract the expression for the muon anomalous magnetic moment ($\mu$AMM) induced by the Leptoquark $S_1$, that is

\begin{align}\label{muAMM}
 a_\mu^{S_1} &= - \frac{ N_c m_\mu }{8 \pi^2 m_{S_1}^2} \Bigl [ m_k \text{Re}( \tilde y^L_{k 2} y^{R*}_{k 2}) ( Q_S G_4(b_k) - Q_{\bar u} G_3 (b_k) ) \nonumber \\
	& -  m_\mu ( | y^{\prime L}_{k 2}|^2 + |y^R_{k 2}|^2 )  (  Q_S G_2 (b_k)  - Q_{\bar u} G_1 (b_k) ) \Bigr ].
\end{align}

In the limit $m_{k}^2 \ll m_{S_1}^2$, the $\mu$AMM can be reduced to 

\begin{equation}
 a_\mu^{S_1} \approx - \frac{N_c m_\mu m_k}{48 \pi^2 m_{S_1}^2} \text{Re} ( y^{\prime L}_{k2} y^{R*}_{k2}) \left [ 4 \log \left ( \frac{m_k^2}{m_{S_1}^2} \right) + 7 \right ]. 
\end{equation}

It should be noted that $\mathcal B (l_i^- \to l_j^- \gamma) $ and $a_\mu^{S_1}$ are considerable enhanced by the masses of the up quarks, mainly the top quark.\\
\\ 
 
\begin{figure}[hbt!]
\centering
\includegraphics[width=15 cm]{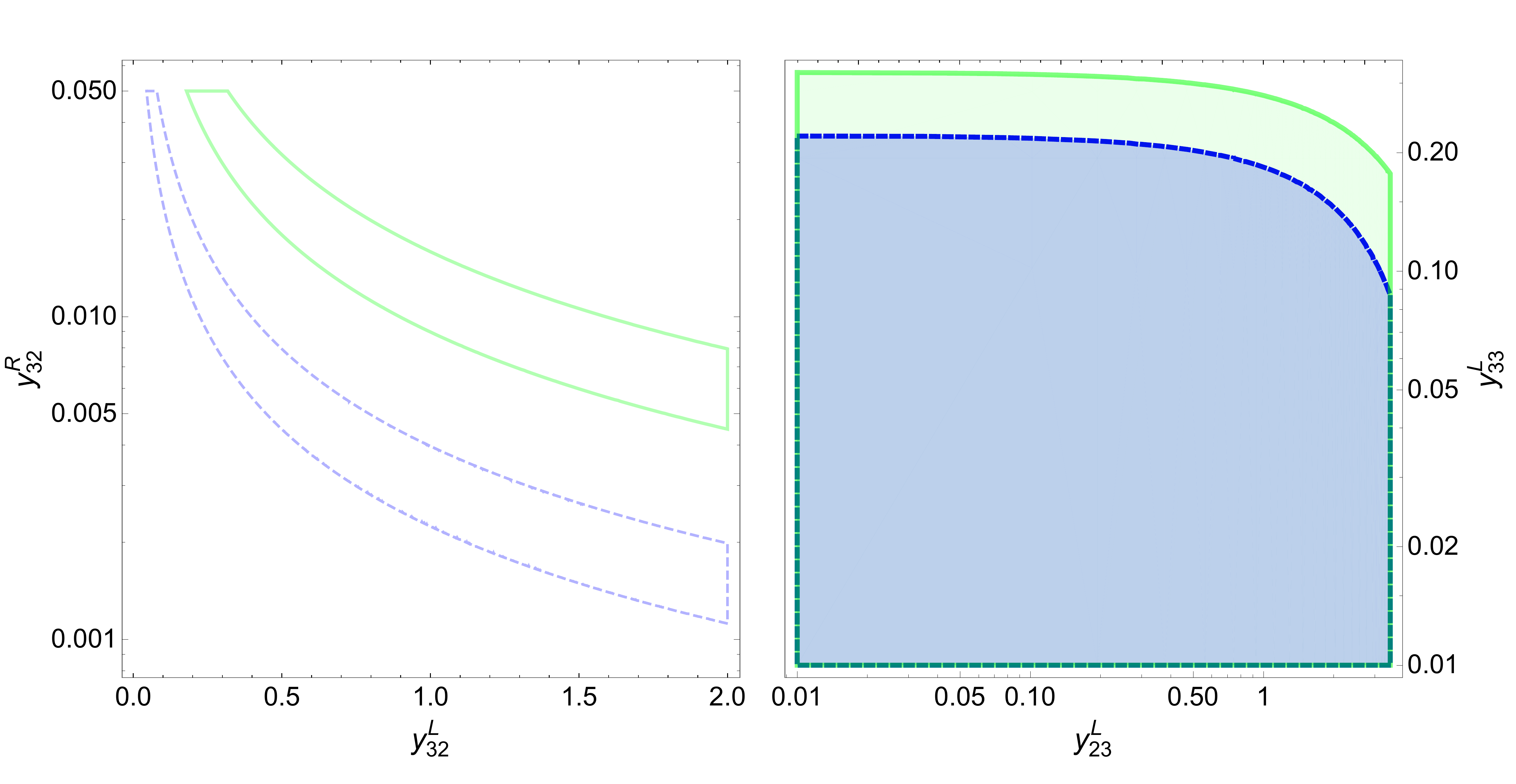}
\caption{Allowed areas with $95 \%$ C.L. of the parameter space for the LQ model assuming that the contribution of $S_1$ is responsible for the discrepancy of the muon anomalous magnetic moment (left plot) and the experimental bound on the $\tau \to \mu \gamma $ decay (right plot) for two values of $m_{S_1}$: $m_{S_1}=1$ TeV (dashed-line boundary) and $m_{S_1} = 3$ TeV (solid-line boundary). }\label{Fig:SpaceParameter1} 
\end{figure}

Now, we implement both processes discussed above and make a preliminary study of the space parameter in order to find the allowed values for the Yukawa LQ couplings, primarily the left-handed ones. To simplify our study we set to zero the LQ couplings $y^L_{1i}\quad (i=1,3)$ and $y^L_{22}$, as they remain largely suppressed by the corresponding elements of the CKM matrix. We also take into account the perturbative limit for the LQ couplings to fermions, which reads as $y^{L, R}_{ij} < \sqrt{4 \pi}$. For the mas of the LQ, we carry out the analysis for both $m_{S_1} = 1$ and 3 TeV. Although one can consider larger values for $m_{S_1}$, it turns out that the  Leptoquark couplings to fermions remain less constrained as the LQ mass  increases. \\ 
\\
To constraints the LQ couplings $y^L_{32}$ and $y^R_{32}$, we assume that $S_1$ is responsible for the discrepancy between the theoretical and the experimental values of $\mu$AMM: $\Delta a_\mu = a_\mu^{\text{Theo.}} - a_\mu^{\text{Exp.}} = 251(59) \times 10^{-11}$, where the experimental data has been recently updated by the Fermilab experiment \cite{Muong-2:2021ojo}. In the left panel of Fig. \ref{Fig:SpaceParameter1} we show the allowed areas by the discrepancy $\Delta a_\mu$ in the plane $y^L_{32}$ vs $y^R_{32}$ for two values of $m_{S_1}$. As expected, for large values of $y^L_{32} \sim \mathcal{O}(1)$ the right-handed coupling $y^R_{32}$ is largely suppressed, being of the order of $\mathcal{O}(10^{-3})$ for $m_{S_1} = 1$ TeV. Conversely, for small values of the left-handed couplings, for example $y^L_{32}\sim .1$, the right-handed interactions can reach large values, now on the order of $O(10^{-2})$.  As a rule, the allowed values for the right-handed coupling can be relaxed up to one order of magnitude as the LQ mass increases to 3 TeV.  This is a comprehensive behaviour because the loop functions of Eq. \eqref{muAMM} are suppressed as soon as the LQ mass increases, so large values for the Yukawa couplings are needed to explain the discrepancy on the $\mu$AMM. Then, in the following analysis we consider the conservative case $y^L_{32} = 1$ since larger values can violate the perturbative limit. \\
\\
Now we consider the experimental limit on $\mathcal{B} (\tau \to \mu \gamma)$ \cite{BaBar:2009hkt} and define restrictions on the parameters $y^L_{i3}\; (i = 2,3)$. Based on the above exploration of $\Delta a_\mu$ we set $y^L_{32} = 1$ and $y^R_{32} = .003\; (.01)$ for $m_{S_1} = 1\; (3)$ TeV. In analogy with the constraints for $y^R_{32}$ we consider the scenario where $y^R_{23} = \epsilon \; y^L_{23}$ with $\epsilon = 10^{-3} \; (10^{-2})$ for $m_{S_1} = 1 \; (3)$ TeV. Therefore, we show in the graph on the right of Fig. \ref{Fig:SpaceParameter1} the allowed areas in the plane $y^L_{23}$ vs $y^L_{33}$ for two values of $m_{S_1}$. We notice that $y^L_{23}$ is not completely suppressed and can take values close to the perturbative limit. As for the LQ coupling to third generation fermions, we manage to find the limit $y^L_{33} < .2$ at $m_{S_1} = 1$ TeV and the allowed area slightly relaxes with increasing the LQ mass. Such restrictions on the LQ couplings may be explained by the fact that in the effective coupling $y^{\prime L}_{33}$ the corresponding parameter $y^L_{23}$ is suppressed by the $V_{23}$ component of the CKM matrix.  It is worth mentioning that we are not taking into account the contribution of the quark charm to the loop, as the primary contribution comes from the top quark. However, the coupling $y^L_{2i}, \; i = 2,3$ is presented since we are working in the down-quark basis.

With this, we have found several constraints on the Left-handed LQ couplings with fermions, however, in order to study the neutrino magnetic moment we also need restrictions on the right-handed LQ couplings with the quark-neutrino pair ($\tilde{y}^{R}_{ij}$). Thereby we also take into account the purely leptonic decay of the $B_c$ meson, the investigation of which will be shown later 

\subsection{ Purely leptonic decay  $B_c^- \to \tau^- \bar \nu$ }

\begin{figure}[hbt!]
\centering
\includegraphics[width=11cm]{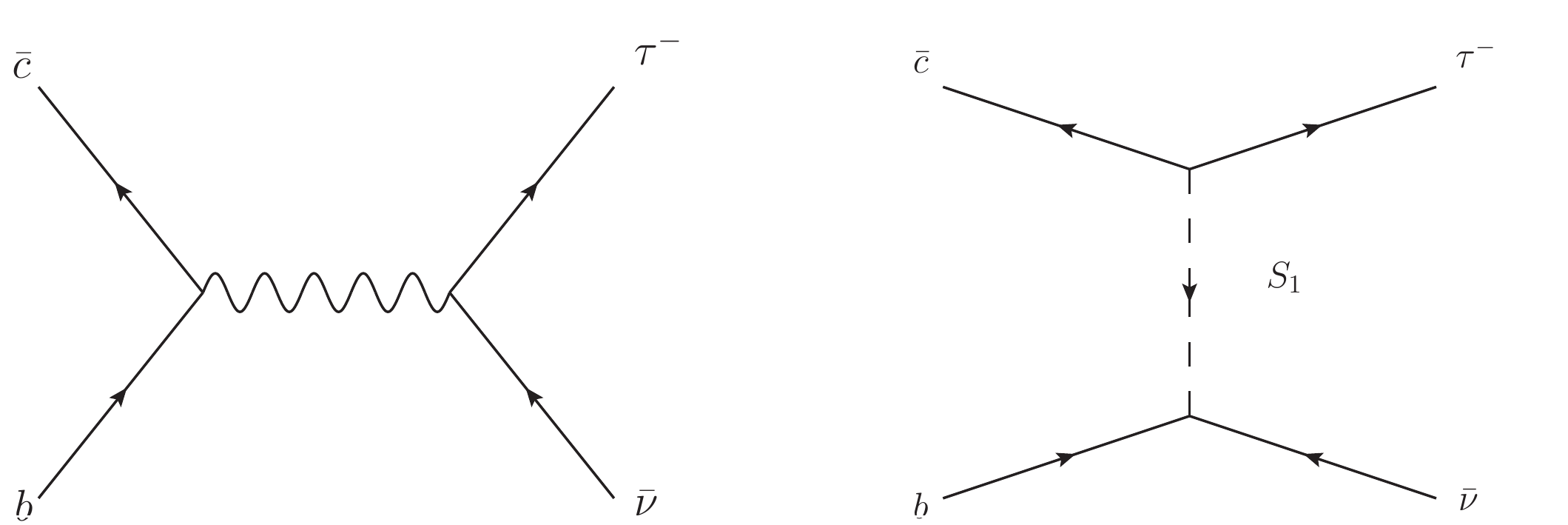}
\caption{Feynman diagrams contributing to $B_c \to \tau^- \bar \nu$ in the Standard model (left diagram) and the new physics contribution of the scalar Leptoquark $S_1$ (right diagram).}\label{Fig:Diagram2} 
\end{figure}

As discussed earlier, the LFV decay $\tau \to \mu \gamma$ and the $\mu$AMM can constraint the left and right-handed Leptoquark couplings $y^L_{ij}$ and $y^R_{ij}$. On the other hand, accurate measurements of the branching ratios of the $B$ meson decays play a critical role in restricting the physics beyond the SM. In this work we also focus on the pure leptonic decays of heavy charge mesons, namely the $B_c$ meson which is a state composed of  bottom and charm quarks. Then, the LQ couplings $y^L_{ij}$ and $\tilde{y}^{R}_{ij}$ have the possibility of being constrained by the experimental limit on the pure leptonic decay $B_c \to \tau^- \bar{\nu}$, which is mediated by the scalar Leptoquark $S_1$ as shown in Fig. \ref{Fig:Diagram2} for the SM and the Leptoquark contribution.

Complementarity, low energy precision flavor observables have the potential to be a test for the direct detection of new particles at high energy. Among the most feasible processes that may indicate the existence of Leptoquarks are the decays of charged pseudoscalar mesons, where the strong helicity suppression of left-handed current operators make this decays quite sensitive to Leptoquarks. Then, the branching ratios are distributed proportionally to $m_l^2$ which means that the decays of the $B$ mesons to light leptons are quite suppressed and currently only bounded from above. Theoretically, the only significant hadronic access for leptonic processes is a decay constant, determined by means of

\begin{equation}
    \langle 0 | \bar c \gamma^\mu \gamma^5 b |B(p)\rangle = i f_{B_c} P^\mu.
\end{equation}
 
Using the equation of motion, one can express pseudoscalar current matrix factor as 

\begin{equation}
    \langle 0|\bar{c} \gamma_5 b|B_c(p)\rangle = -\frac{i f_{B_c}m_{B_c}^2}{m_b(\mu_b) + m_c(\mu_b)},
\end{equation}

where $m_b$ and $m_c$ are the current quark masses given at the scale $\mu_b = m_b$, $f_{B_c} = 0.434$ GeV is the decay constant and $m_{B_c} = 6.275$ GeV is the mass of the $B_c$ meson. The amplitude for the decay $B_c \to \tau \bar{\nu}_j$ mediated by the scalar Leptoquark can be written as follows 

\begin{align}
\mathcal{A}_{B_c \to \tau \nu_j} &= \sqrt{2}G_f f_{B_c}\left\{ \left[ -m_\tau \left(\delta_{lk} V_{bc} + y^{L}_{3\nu} \tilde{y}^L_{23} / \omega \right)- \frac{y^L_{3 j} y^R_{23} m_{B_c}^2}{ \omega (m_b+m_c)} \right] \bar u(p_\tau) P_L v(p_\nu) \right. \nonumber \\
    &+ \left.  \frac{1}{\omega } \left[  y^R_{23} y^{\prime R}_{3\nu} m_\tau   +\frac{y^{\prime R}_{3\nu} \tilde{y}^L_{23} m_{B_c}^2}{ (m_b + m_c) } \right] \bar u(p_\tau) P_R v(p_\nu) \right\},
\end{align}

where $\omega = 4 \sqrt{2} G_f m_{S_1}^2$ with $G_f$ the Fermi constant. By squaring the amplitude and summing over all  spins one obtain the following width decay 

\begin{align}
    \Gamma_{B_c \to \tau \nu_j} &= \frac{ G_f^2 f_{B_c}^2}{8 \pi m_{B_c}^3} (m_\tau^2-m_{B_c}^2)^2 \left( m_\tau^2 |V_{cb}|^2 +\frac{2 m_\tau}{\omega} \text{Re} \sum_{j} \Bigl[ V_{cb}\Bigl( y^{L}_{3j} \tilde{y}^L_{23}  m_\tau \right. \nonumber  \\
    & + \frac{y^L_{3 j} y^R_{23} m_{B_c}^2}{(m_b+m_c)} \Bigr)  \Bigr] + \frac{1}{\omega^2} \sum _j \left| y^{L}_{3j} \tilde{y}^L_{23} m_\tau  + \frac{y^L_{3 j} y^R_{23} m_{B_c}^2}{(m_b+m_c)} \right|^2 \nonumber \\
    &+ \left. \frac{1}{\omega^2} \sum_j \left| y^R_{23} y^{\prime R}_{3 j} m_\tau +\frac{y^{\prime R}_{3 j} \tilde{y}^L_{23} m_p^2}{(m_b + m_c)}\right|^2   \right ).
 \end{align}   

\begin{figure}[hbt!] \label{Fig:SpareParameter2} 
\centering
\includegraphics[width=8cm]{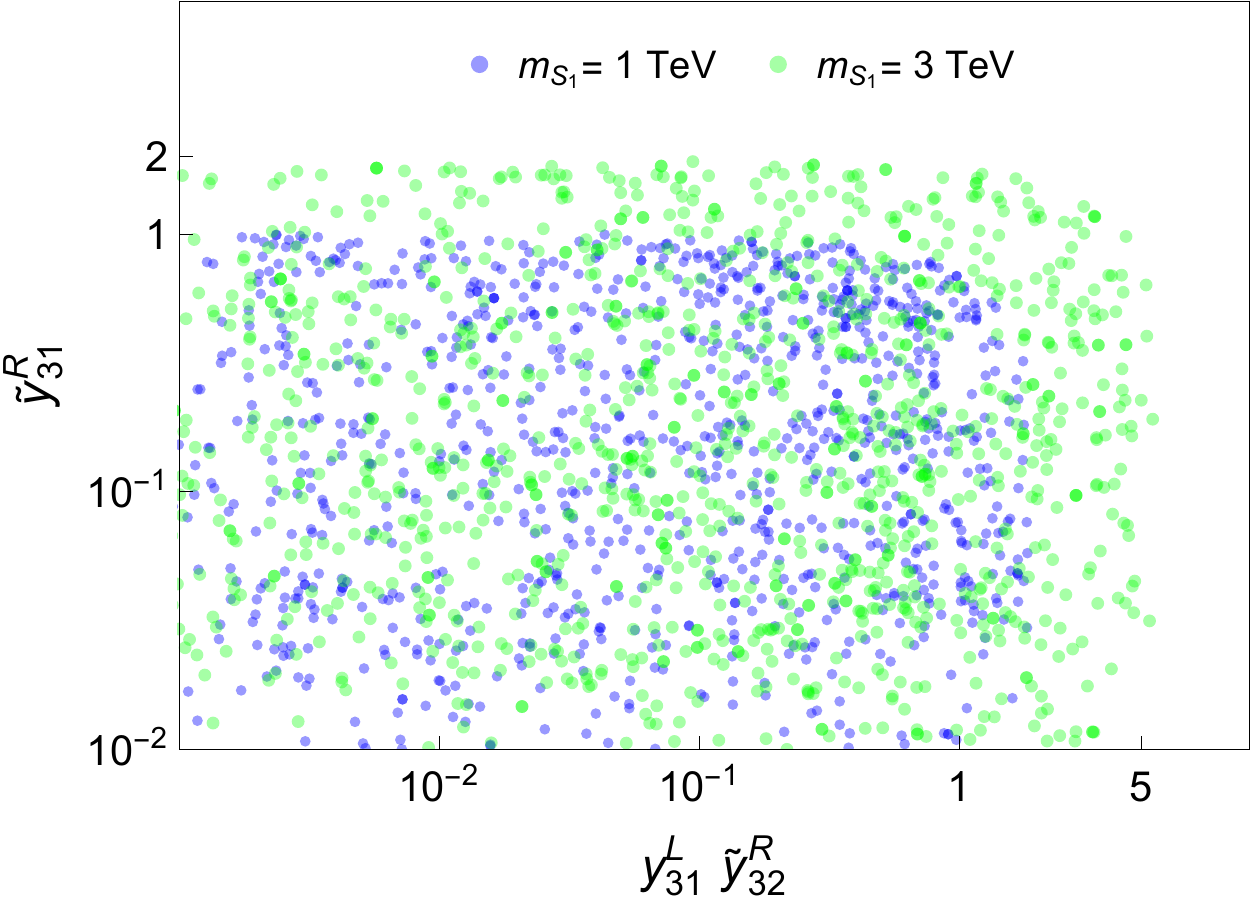}
\caption{Allowed areas with $95\%$ C.L. in the plane $y^L_{31} \tilde{y}^R_{32}$ vs $\tilde{y}^R_{31}$ consistent with the experimental bound on the $B_c \to \tau \bar{\nu}$ decay for $m_{S_1} = 1$ TeV (blue points) and 3 TeV (green points).}
\end{figure}

Now, we use the experimental constraint $\mathcal{B}(B_{c} \to \tau \bar{\nu}) < 10 \%$ \cite{Akeroyd:2017mhr} and perform a scan of $(y^L_{31}, \tilde{y}^{R}_{31}, \tilde{y}^{R}_{32})$ points for two values of the LQ mass. For the other input parameters, we use the values presented in Tab. \ref{tab:Couplings} which are in agreement with the discrepancy $\Delta a_\mu$ and the $\tau \to \mu\gamma$ decay. For the LQ coupling with a neutrino-quark pair of third generation, we consider the conservative value $\tilde{y}^R_{33} = 1$ for $m_{S_1} = 1$ and for $m_{S_1} = 3 $ TeV we set $y^R_{33} = 3$, since smaller values yields to regions out of interest for the analysis of the neutrino magnetic moment. We thus show in Fig. \ref{Fig:SpareParameter2} the allowed area in the plane  $y^L_{31} \tilde{y}^{R}_{32}$ vs $\tilde{y}^{R}_{31}$ for the two scenarios of the LQ mass. Note that the LQ couplings are less restrictive for the LQ masses considered in this paper, for example, for $m_{S_1} = 1$ TeV, the couplings $\tilde{y}^{R}_{32}$ and the product  $y^L_{31} \tilde{y}^{R}_{32}$ can reach values up to $\mathcal O(1)$. The allowed area increases slightly if the LQ mass is now $m_{S_1} = 3$ TeV.

\begin{table}[t]
\begin{center}
\begin{tabular}{| c | c | c | c | c | c | c |}
\hline 
$m_{S_1}$ [TeV]   & $y^L_{32}$   & $y^{R}_{32}$   & $y^L_{23}$   & $y^L_{33}$   & $y^R_{23}$   & $\tilde{y}^{ R}_{33}$\\ 
\hline \hline
$1$  & $1$   & $.003$   & $.8$   & $.15$   & $10^{-3} y^L_{23}$   & $1$    \\
\hline
$3$    &$1$  & $.01$   & $3$   & $.2$   & $10^{-2} y^L_{23}$ &   $3$    \\
\hline
\end{tabular}
\caption{Sample values of the LQ couplings to fermions for two scenarios of $m_{S_1}$ satisfying the constraints of the muon anomalous magnetic dipole moment, the LFV tau decay and the  $B_c \to \tau \bar{\nu}$ decay.}
\label{tab:Couplings}
\end{center}
\end{table}

\section{Numerical analysis of the neutrino magnetic moment}\label{sec:Analysis}

\begin{figure}[hbt!]
\centering
\includegraphics[width=15cm]{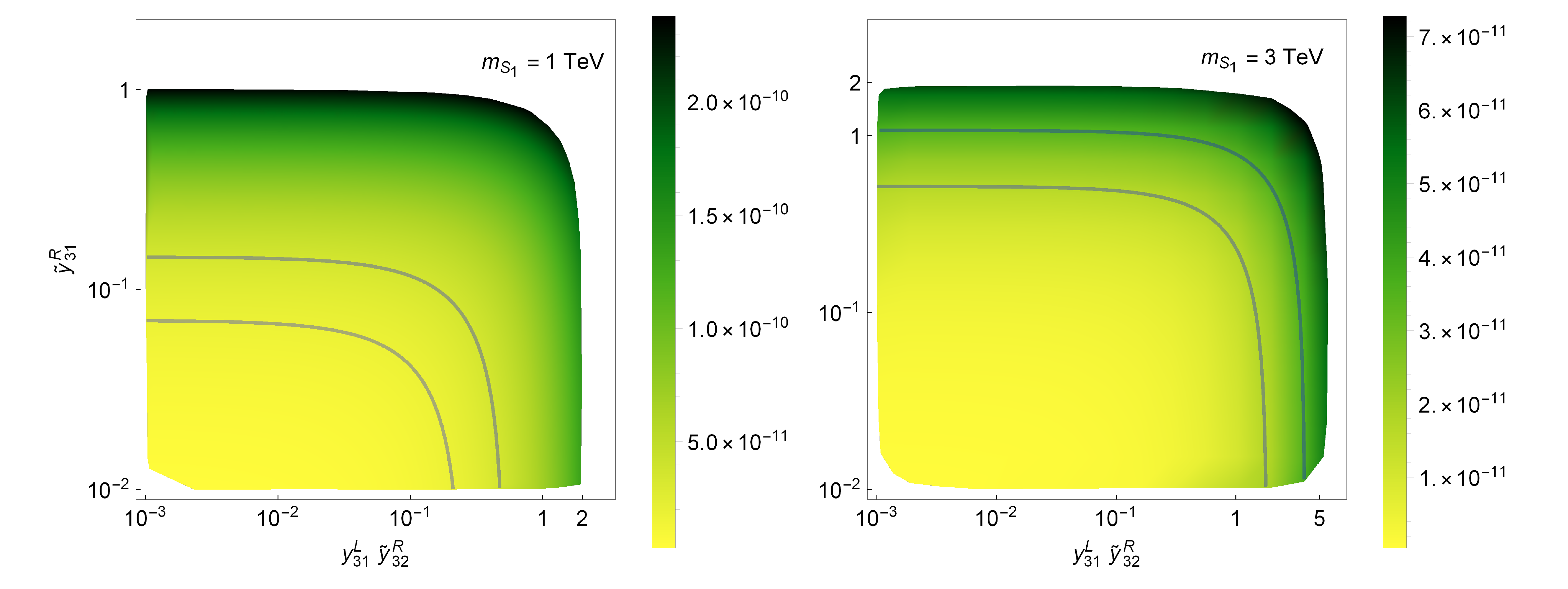}
\caption{Density plots of the transition neutrino magnetic moment $\mu_{\nu_e \nu_\mu}$ for $m_{S_1} = 1$ TeV (left plot) and 3 TeV (right plot) in the plane $(y^L_{31} \tilde{y}^R_{32})$ vs $\tilde{y}^R_{31}$. The solid lines indicate the present XENON1T limit on the neutrino magnetic moment at $90 \%$ confidence level.}\label{Fig:Analysis} 
\end{figure}

Now, we estimate the contribution of the scalar LQ $S_1$ to the neutrino magnetic moment for the allowed  points $(y^L_{31}, \tilde{y}^{R}_{31}, \tilde{y}^{R}_{32})$. Therefore, we show in Fig. \ref{Fig:Analysis} the values of the neutrino magnetic dipole moment in the plane $y^L_{31} \tilde{y}^{R}_{32}$ vs $\tilde{y}^{R}_{31}$  for two values of the LQ mass. For $m_{S_1} = 1$ TeV (left plot) the neutrino magnetic moment reaches its largest value being of the order of $10^{-10}$ for $\tilde{y}^{R}_{31} \sim \mathcal{O}(1)$, however the $\nu$MM  decreases by one order of magnitude when the LQ mass increases up to 3 TeV.   

We have also included the region (bounded by the solid line) where the neutrino transitions magnetic moment can explain the excessive electron recoil events recently reported by the XENON1T experiment. The preferred values of the neutrino transition magnetic moment $\mu_{\nu_e \nu_\mu}$ for the excess observed at the XENON1T experiment at $90 \%$ confidence interval corresponds to $\mu_{\nu_e \nu_\mu} \in (1.65, 3.42) \times 10^{-11} \mu_B$ \cite{Babu:2020ivd}. Then, for $m_{S_1} = 1$ TeV, the values of the LQ couplings needed to explain the XENON data must be of order of $\tilde{y}^{R}_{31} \sim 10^{-1}$ with the product $y^L_{31} \tilde{y}^{R}_{32}$ being in the range $(10^{-3}, 10^{-1})$. However for the LQ coupling product around $3 \times 10^{-1}$, the coupling $\tilde{y}^{R}_{31}$ is allowed to be in the range $(10^{-2}, 10^{-1})$. A similar behaviour is shown for $m_{S_1} = 3$ TeV, where the coupling $\tilde y^{R}_{31}$ can be of the order of $\mathcal{O}(1)$ for LQ coupling product values in the range ($10^{-3},1)$. 

\section{Summary and outlook}\label{sec:Summary}

The neutrino transition magnetic moment was calculated in a simple LQ model, where the SM is augmented by a $SU(2)$ scalar LQ singlet with hypercharge $Y = 1/3$. Then, the scalar LQ in the TeV mass range when couple to both left- and right-handed neutrinos can generate a neutrino magnetic moment, which is enhanced by the quark mass running in the loop. As for the numerical analysis, to obtain bounds on the parameter space we used the experimental constraints on the muon anomalous magnetic moment and the LFV decay of the tau lepton $\tau \to \mu \gamma$ to restrict the left-handed LQ coupling to fermions. We also employ the experimental limit on the pure leptonic decay $B_c \to \tau^- \bar \nu$ to restrict the LQ coupling $\tilde{y}^R_{ij}$. Afterwards the neutrino magnetic moment was evaluated in the allowed region of the parameter space for two values of the LQ mass. In particular, we demonstrate that exist a tiny area where  the model can explain the recently reported excess of electron recoil events by the XENON1T experiment. For the LQ mass of $1\; (3)$ TeV, the area is obtained when the LQ coupling $\tilde y^R_{31}$ is of the order of $10^{-1} \; (1)$ and the coupling product $y^L_{31} \tilde y^R_{32} < 10^{-1} \; (1)$.  

\section*{Acknowledgement}
This work has been supported  by CONACYT-Mexico under grant A1-S-23238.

\printbibliography
\end{document}